# An Efficient and Publicly Verifiable Id-Based Multi-Signcryption Scheme

**Munendra Agrawal, Prashant Kushwah and Sunder Lal**
Department of Mathematics, IBS, Khandari
Agra- 282002 (UP) – INDIA
E-mail: munendra.agrawal1@gmail.com, pra.ibs@gmail.com,
sunder_lal2@rediffmail.com

**Abstract:** Multi-signcryption is used when different senders wants to authenticate a single message without revealing it. This paper proposes a multi signcryption scheme in which no pairing is computed on the signcryption stage and the signatures can be verified publicly.

**Keywords:** Identity based cryptography, Signcryption, Multi-Signcryption, Bilinear pairing.

**1. Introduction:** Secure message transmission over an insecure channel, require both confidentiality and authenticity, which may be achieved through 'signature then encryption' approach. However, in 1997 Zheng [24] proposed a cryptographic primitive "Signcryption" which achieves both confidentiality and authenticity in a single logical step with much lower computational cost than signature then encryption approach. Beak et al. [3] gave formal security model for signcryption scheme and provided security proof for Zheng's scheme in random oracle model.

1984, Shamir [20] introduced the concept of identity based cryptography and gave the first identity based signature scheme. The idea of identity based cryptography is to enable a user to use any arbitrary string that uniquely identifies him as his public key. Identity based cryptography serves as an efficient alternative to Public Key Infrastructure (PKI) based system. In 2001, Boneh and Franklin [5] gave the first identity based encryption scheme and in 2003 Malone-Lee [14] gave the first identity based signcryption scheme. He also considered security notions of signcryption in identity based setting. Since then many signcryption schemes have been proposed [1, 7, 8, 9, 13, 16].

In 1989, Boyed [6] defined multi signature scheme the scenario where more than one user authenticate a single message in such a way that a verifier verify only a single compact signature on that message. An identity based version of multi-signature was given by Gentry et al. [11] in 2006. Mitomi et al. [15] include confidentiality in multi-signature by signcrypting the message. Zhang et al. [22] gave an identity based multi-signcryption scheme. However, S. Deva et al. [18] find some flaw and fix them. Recently Zhang et al. [23] came up with a new multi-signcryption scheme in identity based setting which to the best our knowledge is most efficient scheme till date.

The concept of multi-receiver setting was first given by Bellare et al. [4] for public key encryption where, there are $n$ receivers numbered by $1,\ldots,n$ and each of them generates for itself a private key and public key pair denoted by ($sk_i$, $pk_i$). A sender encrypts a message $m$ using $pk_i$ to obtain $C_i$ for $i = 1,\ldots,n$ and then sends ($C_1,\ldots,C_n$) as a ciphertext. Upon receiving the ciphertext, receiver $i$ extract $C_i$ and decrypt it using $sk_i$. Beak et al. [2] formalized identity based encryption to the multiple receivers setting. Duan and Cao [10] consider the situation where there is not only multiple receiver but also multiple senders. As an example, consider that there are several managers, each of whom wants to securely broadcast an e-mail to the employees of the company independently. Once an employee receives several ciphertexts from different managers, an issue of message authentication will arise. In such case confidentiality and authenticity required simultaneously. Motivated by this Duan and Cao [10] gave the first multi-receiver identity based signcryption scheme. Later on some more multi-receiver identity based signcryption schemes were proposed [12, 17, 19, 21].

In this paper, we propose an identity based multi- signcryption scheme which is more efficient then the schemes S. Deva et. al [18] and Zhang et. al [23]. Our scheme needs no pairing computation on the signcryption stage. The scheme also possesses public verifiability for signature. We also convert our proposed scheme for multiple receivers, which we believe is the first multi-signcryption scheme for multiple receivers.

## 2. Preliminaries:

Let $\mathbb{G}_1$ be an additive group and $\mathbb{G}_2$ be a multiplicative group, of the same prime order $q$. A function $e: \mathbb{G}_1 \times \mathbb{G}_1 \to \mathbb{G}_2$ is called a **bilinear pairing** if it satisfies the following properties:

(i) $\forall P, Q \in \mathbb{G}_1, \forall a, b \in \mathbb{Z}_q^*, e(aP, bQ) = e(P, Q)^{ab}$

(ii) For any point $P \in \mathbb{G}_1, e(P, Q) = 1$ for all $Q \in \mathbb{G}_1$ if only if $P = \mathcal{O}$, the identity of $\mathbb{G}_1$

(iii) There exists an efficient algorithm to compute $e(P, Q) \, \forall P, Q \in \mathbb{G}_1$.

Given $(P, aP, bP) \in \mathbb{G}_1$ for unknown $a, b \in \mathbb{Z}^*$, the Computational Diffie-Hellman Problem (**CDH Problem**) in $\mathbb{G}_1$ is to compute $abP$.

Given two groups $\mathbb{G}_1$ and $\mathbb{G}_2$ of the same prime order $q$, a bilinear map $e: \mathbb{G}_1 \times \mathbb{G}_1 \to \mathbb{G}_2$, a generator $P$ of $\mathbb{G}_1$, three elements $aP, bP, cP$ of $\mathbb{G}_1$ and an element $H \in \mathbb{G}_2$, the Decisional Bilinear Diffie-Hellman Problem (**DBDH Problem**) is to decide whether $H = e(P, P)^{abc}$.

Before giving the proposed multi-signcryption scheme, first we formalize **Identity based Multi-Signcryption**

An Identity based Multi- Signcryption scheme consists of the following algorithms:

- **Setup:** Given a security parameter k, the Private Key Generator (PKG) chooses a secret value randomly and generates master secret key *msk* and the public parameters *params* of the system.

- **Key Extract:** Given a user identity $ID \in \{0,1\}^*$, the PKG computes the corresponding private key $S$ and transmits it to the user in a secure way.

- **Signcrypt:** Different users with identities $L = \{ID_1, ..., ID_n\}$ run this algorithm to signcrypt a message $m$ to the receiver's identity $ID_B$ to obtain signcryption $\sigma$.

- **Unsigncrypt:** The receiver B with identity $ID_B$ and private key $S_B$ runs this algorithm to obtain plaintext $m$, if $\sigma$ is a valid signcryption from L to identity $ID_B$ otherwise return $\perp$.

For consistency, we require that if

$$\sigma = \text{signcrypt}(m, L = \{ID_1, ..., ID_n\}, S_1, ..., S_n, ID_B), \text{then}$$
$$m = \text{unsigncrypt}(\sigma, L = \{ID_1, ..., ID_n\}, ID_B, S_B).$$

### 3. The proposed scheme:

**Setup:** Given $k$ is the security parameter, the PKG chooses the system parameter that includes two groups $\mathbb{G}_1, \mathbb{G}_2$ of same prime order $q$, a bilinear map $e: \mathbb{G}_1 \times \mathbb{G}_1 \to \mathbb{G}_2$, a generator $P \in \mathbb{G}_1$, randomly chosen $s \in_R \mathbb{Z}_q$, $R \in_R \mathbb{G}_1 (R \neq P, \mathcal{O})$ and computes $P_{pub} = sP \in \mathbb{G}_1$ and $\theta = e(P_{pub}, R)$. The PKG also chooses cryptographic hash functions $H_0: \{0,1\}^* \to \mathbb{G}_1$, $H_1: \mathbb{G}_2 \to \{0,1\}^l$, $H_2: \{0,1\}^l \times \mathbb{G}_1 \times \mathbb{G}_1 \to \mathbb{Z}_q^*$ where $l$ is the length of plaintext and ciphertext.

The system public parameters are

$$\text{params} = \langle q, \mathbb{G}_1, \mathbb{G}_2, e, l, P, P_{pub}, \theta, R, H_0, H_1, H_2 \rangle$$

**Key Extract:** Given a user identity $ID \in \{0,1\}^*$, PKG compute public key $Q_{ID} = H_0(ID)$ and private key $S_{ID} = sQ_{ID}$.

**Signcrypt:** Given a message $m \in \{0,1\}^l$, a receiver's identity $ID_B$ and $n$ senders' identities $L = \{ID_1, ..., ID_n\}$, each user $ID_i$ executes the following steps

(i) Randomly chooses $x_i \in \mathbb{Z}_q^*$ and computes $X_i = x_i P$, $Y_i = \theta^{x_i}$ and $U_i = x_i(R + Q_B)$.

(ii) Sends $(X_i, Y_i, U_i)$ to other signers through a secure channel.

(iii) After receiving from the other signers $(X_j, Y_j, U_j)$, user $ID_i$ computes

(a) $X = \sum_{i=1}^{n} X_i$, $Y = \prod_{i=1}^{n} Y_i$, $Q = \sum_{i=1}^{n} Q_i$ and $U = \sum_{i=1}^{n} U_i$.

(b) $c = H_1(Y) \oplus m$.

(c) $h = H_2(c, X, U)$.

(d) $Z_i = hS_i + x_i Q$

(iv) Each user sends $Z_i$ to other users. Every user computes $Z$ and outputs ciphertext $\sigma$, where $Z = \sum_{i=1}^{n} Z_i$ and $\sigma = \langle c, X, Z, U, L \rangle$.

**Unsigncrypt:** To unsigncrypt the ciphertext $\sigma = \langle c, X, Z, U, L \rangle$, the receiver with identity $ID_B$ computes

(i) $h = H_2(c, X, U)$, and accepts iff

(ii) $e(P, Z) = e(X + hP_{pub}, Q)$.

It then computes

(iii) $Y' = e(P_{pub}, U)e(X, S_B)^{-1}$, and recovers

(iv) $m = c \oplus H_1(Y')$

**4. Security :**

**(i) Confidentiality:** Without knowing the secrete key of receiver, no one can compute $Y = \prod_{i=1}^{n} Y_i = \theta^{(x_1 + \ldots + x_n)} = e(P_{pub}, R)^{(x_1 + \ldots + x_n)}$. It is only the specific receiver who can compute the actual value of Y using secrete key as

$$Y' = e(P_{pub}, U)e(X, S_B)^{-1} = e(P_{pub}, \sum_{i=1}^{n} U_i)e(X, S_B)^{-1}$$

$$= e(P_{pub}, \sum_{i=1}^{n} x_i(R + Q_B))e(X, S_B)^{-1}$$

$$= e(P_{pub}, \sum_{i=1}^{n} x_i R)e(P_{pub} + \sum_{i=1}^{n} x_i Q_B)e(X, S_B)^{-1}$$

$$= e(P_{pub}, (x_1 + \ldots + x_n)R)e(sP, (x_1 + \ldots + x_n)Q_B)e(X, S_B)^{-1}$$

$$= e(P_{pub}, R)^{(x_1 + \ldots + x_n)} e((x_1 + \ldots + x_n)P, sQ_B)e(X, S_B)^{-1}$$

$$= \theta^{(x_1 + \ldots + x_n)} e(X, S_B)e(X, S_B)^{-1}$$

$$= \theta^{(x_1 + \ldots + x_n)} = Y.$$

**(ii) Public Verifiability:** Any one who has access to the signcryptext can verify the signature on the ciphertext which it contains. First the verifier computes $h = H_2(c, X, U)$ and $Q = \sum_{i=1}^{n} Q_i$, then checks

$$e(P, Z) = e(P, \sum_{i=1}^{n}(hS_i + x_iQ)) = e(P, \sum_{i=1}^{n} hS_i)e(P, \sum_{i=1}^{n} x_iQ)$$

$$= e(P, \sum_{i=1}^{n} hsQ_i)e(P, \sum_{i=1}^{n} x_iQ)$$

$$= e(P, hs\sum_{i=1}^{n} Q_i)e(P, (x_1 + \ldots + x_n)Q)$$

$$= e(P, hsQ)e((x_1 + \ldots + x_n)P, Q)$$

$$= e(hsP, Q)e(X, Q)$$

$$= e(hP_{pub}, Q)e(X, Q)$$

$$= e(X + hP_{pub}, Q).$$

**(iii) Unforgeability:** Signcryptext is generated using the secret key $S_i$ of each of the signers. Thus no one, not even the one among signers can generate a valid signcryptext without knowing the secret key of **all** the signers.

**5. Efficiency Comparison:** We compare efficiency of our scheme with existing schemes [18, 23]. We consider the costly operations which include scalar multiplications in $\mathbb{G}_1$ ($\mathbb{G}_1$ Mul), exponentiations in $\mathbb{G}_2$ ($\mathbb{G}_2$ Exp) and pairing operations (Pairing) as shown below

|  | Signcrypt | | | Unsigncrypt | | |
| --- | --- | --- | --- | --- | --- | --- |
|  | $\mathbb{G}_1$ Mul | $\mathbb{G}_2$ Exp | Pairing | $\mathbb{G}_1$ Mul | $\mathbb{G}_2$ Exp | Pairing |
| S. Deva et al. [18] | 3n | n | n | 0 | 1 | 4 |
| Zhang et al. [23] | 4n | 0 | n | 0 | 1 | 4 |
| Proposed Scheme | 4n | n | 0 | 1 | 0 | 4 |

**Table - 1**

**6. Remarks:**

**(i)** In the proposed schemes, we use the concept of Duan and Cao [10] which they proposed to construct an Identity based Multi-receiver signcryption scheme.

**(ii)** One of the important advantages of the proposed scheme is that no pairing computation is needed for signcryption. This makes the scheme quite efficient.

**(iii)** To achieve efficiency in Zhang et al. scheme [23], only one signer will compute the signcryptext but in our scheme every user can generates own copy of signcryptext.

**(iv)** The proposed scheme is publicly verifiable. Any one who can access to the signcryptext can verify the signature on ciphertext $c$. Thus the proposed scheme is more applicable when signing a joint confidential contract between two or more organizations. Any one can verify the authenticity of the contract without getting any knowledge of it, however, only the authority can read the contract.

**7. Identity Based Multi-Signcryption Scheme for Multiple Receiver:**

An Identity based Multi- Signcryption scheme for Multi-Receivers consists of the following algorithms:

- **Setup:** Given a security parameter k, the Private Key Generator (PKG) chooses a secret value randomly and generates master secret key *msk* and the public parameters *params* of the system.

- **Key Extract:** Given a user identity $ID \in \{0,1\}^*$, the PKG computes the corresponding private key $S$ and transmits it to the user in a secure way.

- **Signcrypt:** Any $n$ users $L = \{ID_1,...,ID_n\}$ run this algorithm to signcrypt a message $m$ to any $n'$ receiver's with identities $L^* = \{ID'_1,...,ID'_{n'}\}$, and to obtain signcryption $\sigma$.

- **Unsigncrypt:** Each receiver with identity $ID'_j$ and private key $S'_j$ runs this algorithm to obtain plain text $m$ if $\sigma$ is a valid signcryption from $L$ to identity $ID'_j$ otherwise return $\perp$.

For consistency, we require that if

$$\sigma = \text{signcrypt}(m, L = \{ID_1, ..., ID_n\}, S_1, ..., S_n, L^* = \{ID'_1, ..., ID'_{n'}\}), \text{ then}$$
$$m = \text{unsigncrypt}(\sigma, L = \{ID_1, ..., Id_n\}, L^* = \{ID'_1, ..., ID'_{n'}\}, S'_1, ..., S'_{n'}).$$

**The proposed scheme:**

**Setup:** Given $k$ is the security parameter, the PKG chooses the system parameter that includes two groups $\mathbb{G}_1, \mathbb{G}_2$ of same prime order $q$, a bilinear map $e: \mathbb{G}_1 \times \mathbb{G}_1 \to \mathbb{G}_2$, a generator $P \in \mathbb{G}_1$, randomly chosen $s \in_R \mathbb{Z}_q$, $R \in_R \mathbb{G}_1 (R \neq P, \mathcal{O})$ and computes $P_{pub} = sP \in \mathbb{G}_1$ and $\theta = e(P_{pub}, R)$. The PKG also choose cryptographic hash functions $H_0: \{0,1\}^* \to \mathbb{G}_1$, $H_1: \mathbb{G}_2 \to \{0,1\}^l$, $H_2: \{0,1\}^* \to \mathbb{Z}_q^*$ where $l$ is the length of plaintext and ciphertext.

The system public parameters are

$$\text{params} = \langle q, \mathbb{G}_1, \mathbb{G}_2, e, l, P, P_{pub}, \theta, R, H_0, H_1, H_2 \rangle$$

**Key Extract:** Given a user identity $ID \in \{0,1\}^*$ then PKG compute public key $Q_{ID} = H_0(ID)$ and private key $S_{ID} = sQ_{ID}$.

**Signcrypt:** Given a message $m \in \{0,1\}^l$, $n'$ receiver's identity $L^* = \{ID'_1, ..., ID'_{n'}\}$ and $n$ senders' identities $L = \{ID_1, ..., ID_n\}$, each user $ID_i$ execute the following steps

(i) Randomly chooses $x_i \in \mathbb{Z}_q^*$ and computes $X_i = x_i P$, $Y_i = \theta^{x_i}$ and $U_{i,j} = x_i(R + Q'_j)$ for $j = 1, ..., n'$.

(ii) Sends $(X_i, Y_i, U_{i,1}, U_{i,2}, ..., U_{i,n'})$ to other signers through a secure channel.

(iii) After receiving from the other signers $(X_i, Y_i, U_{i,1}, U_{i,2}, ..., U_{i,n'})$, user $ID_i$ computes

(a) $X = \sum_{i=1}^{n} X_i$, $Y = \prod_{i=1}^{n} Y_i$, $Q = \sum_{i=1}^{n} Q_i$ and $U_1 = \sum_{i=1}^{n} U_{i,1}$, $U_2 = \sum_{i=1}^{n} U_{i,2}, ..., U_{n'} = \sum_{i=1}^{n} U_{i,n'}$.

(b) $c = H_1(Y) \oplus m$.

(c) $h = H_2(c, X, U_1, U_2, ..., U_{n'})$.

(d) $Z_i = hS_i + x_i Q$.

(iv) Each user sends $Z_i$ to other users. Every user computes $Z$ and outputs ciphertext $\sigma$, where $Z = \sum_{i=1}^{n} Z_i$ and $\sigma = \langle c, X, Z, U_1, U_2, ..., U_{n'}, L, L^* \rangle$.

**Unsigncrypt:** To unsigncrypt the ciphertext $\sigma = \langle c, X, Z, U_1, U_2, ..., U_{n'}, L, L^* \rangle$, the receiver with identity $ID'_j$ computes

(i) $h = H_2(c, X, U_1, U_2, ..., U_{n'})$, and accepts iff

(ii) Accept if $e(P, Z) = e(X + hP_{pub}, Q)$.

It then extract $U_j$ from $\sigma$ and computes

(iii) $Y' = e(P_{pub}, U_j) e(X, S'_j)^{-1}$, and recovers

(iv) $m = c \oplus H_1(Y')$

## 8. Security :

**(i) Confidentiality:** Without knowing the secrete key of receiver, no one can compute $Y = \prod_{i=1}^{n} Y_i = \theta^{(x_1+...+x_n)} = e(P_{pub}, R)^{(x_1+...+x_n)}$. It is only the specific receiver who can compute the actual value of Y using secrete key as

$$Y' = e(P_{pub}, U_j) e(X, S'_j)^{-1} = e(P_{pub}, \sum_{i=1}^{n} U_{i,j}) e(X, S'_j)^{-1}$$

$$= e(P_{pub}, \sum_{i=1}^{n} x_i (R + Q'_j)) e(X, S'_j)^{-1}$$

$$= e(P_{pub}, \sum_{i=1}^{n} x_i R) e(P_{pub} + \sum_{i=1}^{n} x_i Q'_j) e(X, S'_j)^{-1}$$

$$= e(P_{pub}, (x_1 + ... + x_n)R) e(sP, (x_1 + ... + x_n)Q'_j) e(X, S'_j)^{-1}$$

$$= e(P_{pub}, R)^{(x_1+...+x_n)} e((x_1 + ... + x_n)P, sQ'_j) e(X, S'_j)^{-1}$$

$$= \theta^{(x_1+...+x_n)} e(X, S'_j) e(X, S'_j)^{-1}$$

$$= \theta^{(x_1+...+x_n)} = Y.$$

**(ii) Public Verifiability:** Any one who has access to the signcryptext can verify the signature on the ciphertext which it contains. First the verifier computes $h = H_2(c, X, U_1, U_2, ..., U_{n'})$, $Q = \sum_{i=1}^{n} Q_i$, then checks

$$e(P, Z) = e(P, \sum_{i=1}^{n} (hS_i + x_i Q)) = e(P, \sum_{i=1}^{n} hS_i) e(P, \sum_{i=1}^{n} x_i Q)$$

$$= e(P, \sum_{i=1}^{n} hsQ_i) e(P, \sum_{i=1}^{n} x_i Q)$$

$$= e(P, hs \sum_{i=1}^{n} Q_i) e(P, (x_1 + ... + x_n)Q)$$

$$= e(P, hsQ) e((x_1 + ... + x_n)P, Q)$$

$$= e(hsP, Q) e(X, Q)$$

$$= e(hP_{pub}, Q) e(X, Q)$$

$$= e(X + hP_{pub}, Q).$$

**(iii) Unforgeability:** Signcryptext is generated using the secret key $S_i$ of each the signers. Thus no one, not even the one among signers can generate a valid signcryptext without knowing the secret key of **all** the signers.

**9. Conclusion:** We have proposed an efficient Identity Based Multi-Signcryption Scheme. We discuss its confidentiality, unforgeability and public verifiability in heuristic way and compare it with two existing Id-based multi-signcryption schemes. We also extend the proposed scheme to multi-signcryption scheme for multiple receivers.